A low-frequency superconductor oscillator with a $10^{10}$–quality factor.

O. F. Schilling

Departamento de Física, Universidade Federal de Santa Catarina, Campus, Trindade, 88040-900, Florianópolis, SC. Brazil.

Abstract.

This letter describes a very simple electromechanical oscillator, consisting of a strong-pinning Nb-Ti superconductor loop subjected to static magnetic fields. A detailed calculation of the losses occurring during its low-frequency oscillations is carried out. The conclusion is that the quality factor for such oscillator might reach the $10^{10}$-$10^{11}$ range, something comparable only to the best optical microcavities. Such device might permit the measurement of variations in static forces with the same precision a superconducting quantum interference device ( SQUID) measures variations of magnetic field, providing a new tool for probing minute variations of gravitational field, for instance.

1. Introduction.

Oscillators are widely used in the range of frequencies from kHz to tens of MHz, in applications like TV transmission and reception, dielectric and magnetic heating, or in electronic instruments for timing and testing purposes. Crystal oscillators with quality factors ($Q$) in the range of $10^5$ or greater are adopted when precise resonant effects are required. In recent years the development of optical devices has led to even more strict requirements on stability. Optical microcavities have been developed with $Q$ factors on the order of $10^9$, and are used in the control of laser-emitted spectra for transmission of data through optical fibers, for instance[1]. In all these cases the high stability is associated with very high frequencies of operation. We have recently developed the project of a simple electromechanical oscillator that would still be extremely stable, even working at frequencies well below the kHz range[2]. It is the objective of this letter to stress some unique properties of such oscillating system, to show theoretically its extreme stability, and to highlight a particularly important application for it on the measurement of gravitational fields. The layout of the oscillator is depicted in Figure 1 [3]. Under certain experimental conditions a type-II superconductor square loop of side length $l$, mass $m$, and self-inductance $L$ would levitate in the upright position while subjected to the uniform distribution of fields $B_1$ and $B_2$. The length of the lower side of the loop subjected to $B_2$ is designated as $a$ ($< l$). According to the theory in [2], in the "absence" of losses the loop would perform quasi-harmonic oscillations of frequency $\Omega = B_0 a/(mL)^{1/2}$, with $B_0 \equiv B_1 - B_2$. The amplitude of the oscillations is $x_0 = g/\Omega^2$, and the maximum speed of the oscillations is $v_0 = g/\Omega$, where $g$ stands for the gravity acceleration ( assuming $F = mg$ in [2]). Such mechanical oscillations are accompanied by a rectified current containing an alternating component of same frequency, and amplitude $i_0 = mg/(aB_0)$. We note that displacement, speed, and the magnitude of the induced current are all directly proportional to the static force, and in the present example, proportional to $g$.

The original work discussed quite extensively how different sources of losses might be avoided by the proper design of the experiment. In particular, the loop should oscillate in a vacuum and electrically insulating magnets should be used, to avoid friction and eddy current losses induced in nearby metallic parts. Losses in the superconductor itself are practically eliminated by working well below $T_c$ and by keeping $\Omega$ well below the MHz range. A strong-pinning material should be used to make the wire, to avoid flux-creep and hysteresis losses, as discussed below. Hysteresis is by far the most important source of energy dissipation in strong-pinning superconductors carrying alternating currents at low frequencies, under external magnetic fields. The key for eliminating hysteresis is not allowing the flux-lines (FL) in the wire to move irreversibly between pinning centers. Such irreversible displacements of the FL are associated with the establishment of a critical-state flux profile in the specimen. We recall some results of Campbell's work on this subject[4,5] quoted in[2]. Campbell demonstrated that if a type-II superconductor sample is subjected to a static external field $H$ superimposed to a small ripple field $h$ of frequency $\omega$, a critical-state flux profile will be established from the surface only if $b = \mu_0 h$ is greater than $\delta \equiv (\mu_0 J_c B d_o)^{1/2}$, where $J_c$ is the critical current density of the material, $B \approx \mu_0 H$, and $d_o$ is the size of the potential energy well that arrests a flux line ($d_o$ is measured in [5] and is on the order of 2 to 6 nm). For lower ripple fields the FL displacements are reversible and no losses were measured. We call this the *reversible limit*. In such conditions the ripple field and the corresponding induced current will be restricted to a surface layer of thickness $\lambda_{eff} = (B d_o/(\mu_0 J_c))^{1/2}$, which is on the order of micrometers for strong-pinning materials and $B$ on the order of Tesla. $\lambda_{eff}$ is analogous to the penetration depth of type-I superconductors, but much greater in magnitude. Therefore, to stay within the reversible limit the current $i$ flowing in a square loop made with a cylindrical superconductor wire of radius $r$ should not generate a field greater than $\delta$ at any point on the surface of the wire, and thus $i$ must be restricted by the criterion [2]

$$\gamma\mu_0 i/(2\pi r) \leq (\mu_0 J_c B d_o)^{1/2} \qquad (1)$$

Here $\gamma \approx 1.75$ corrects for the concentration of field at the inner corners of the loop as compared with the field $\mu_0 i/(2\pi r)$ produced by an infinite wire of same radius( something easily demonstrated by numerically calculating the fields produced by the current on the corners, by means of Biot and Savart's law; the correction factor $\gamma$ is quite insensitive to the wire radius/side length ratio of the loop).

Campbell's theoretical model completely neglects energy dissipation associated with the FL oscillations inside the pinning wells. However, the motion of the FL is accompanied by dragging of the normal electrons in their centers, so that a viscous force term proportional to the velocity of the FL should be included in the theory[6]. In fact, Campbell´s description of the reversible penetration of flux into a planar vacuum/superconductor interface can be deduced from the ordinary treatment of eddy currents in metals[7], provided one assumes an imaginary resistivity $\rho_C = (i\omega\eta/k)(H/H_{c2})\rho_n$, which takes the viscosity effects partially into account. Here the notation of [6] is adopted: $\eta = \phi_0\mu_0 H_{c2}/\rho_n$ is the flux-flow viscosity, $k = J_c\phi_0/d_0$ is the elastic force constant of the pinning interaction, $\rho_n$ is the normal-state resistivity, $H_{c2}$ is the upper critical field, and $\phi_0$ is the flux quantum. With such expression for $\rho_C$ Campbell's effective penetration depth $\lambda_{eff}$ may be obtained at once from the formula for the eddy currents skin depth[7], as $\lambda_{eff} = (\mu_0 \omega / |\rho_C|)^{-1/2}$. For this purely imaginary resistivity the phasors **E** and **J**= **E**/$\rho_C$ become 90º out of phase and thus no power $P = \text{Re} \langle \mathbf{E^*J} \rangle$ related to flux motion is dissipated within this level of approximation.

The more precise treatment of Gittleman and Rosenblum[6] results in a resistivity $\rho_{GR} = (i\omega\eta / (i\omega\eta + k))(H/H_{c2})\rho_n$. There will be losses associated with the real part of $\rho_{GR}$. Since the ratio $\eta/k$ is usually much smaller than $10^{-6}$ these losses become extremely small, but not null at low frequencies. Gittleman and Rosenblum's treatment of viscous-flow losses associated with the motion of FL at low currents can be directly applied to the evaluation of the quality factor $Q$ for the superconductor electromechanical oscillator introduced in [2], since it provides a method for calculating the power losses due to vibrations of pinned FL. We take as a realistic example the losses that would be associated with the oscillations of a loop made with a Nb-48 Wt % Ti alloy described in detail in [8,9]. The physical properties relevant for the calculations were all carefully determined for this material, so that a quite precise application of the theory is possible.

2. Quality factor calculation for a strong-pinning Nb-Ti loop.

Meingast and collaborators[8,9] performed a very detailed study of the physical and microstructural properties of a Nb-48 Wt. % Ti alloy containing a homogeneous distribution of α-Ti precipitates. In the present calculations we take their data for the particular case of a wire drawn to a 0.645 mm diameter. The upper critical field $\mu_0 H_{c2}$= 11.5 T at 4.2 K; the normal state resistivity at the critical temperature (9.5 K) is $\rho_n$= 7x10$^{-7}$ ohm.m; the critical current density $J_c$ = 3x10$^9$ A/m$^2$ at 4.2 K, under a magnetic field of 0.6 Tesla. It will be assumed that the loop is a square of side length $l$= 5 cm, and $a$=3 cm in Figure 1. Since the average density of this Nb-Ti alloy is 6.6 g/cm$^3$, the loop mass would be $m$= 0.431 g. The self-inductance of the loop $L$ can be calculated from a specific formula, so that $L$= 1.5x10$^{-7}$ H. The static fields produced by the magnets will be assumed as $B_1$= 0.6 T and $B_2$= 0.3 T, so that $B_0$= 0.3 T. In order that hysteresis losses be avoided one of the conditions to be met is that the static fields acting upon each point of the wires must remain constant to a precision of about 10% of the ripple field produced by the currents in the loop, irrespective of the loop vertical displacements. From the value for δ ( see below) this tolerance range should be on the order of 4x10$^{-4}$ T. All the parameters needed by the theory can be calculated from these data. We obtain Ω= 1119 rad/s, corresponding to a low frequency of oscillation, $f$ = 178 Hz. The amplitude of oscillation is $x_0$= 7.8 μm, which is quite small. From relation (1) it is possible to calculate the threshold ripple field $\delta$, which is $3.4 \times 10^{-3}$ T, for $B = B_1 = 0.6$ T and $d_0 = 5$ nm in (1) ($d_0$ should be similar to the coherence length measured for the alloy at 4.2 K, $\xi = 5$ nm), which results in the threshold current $i_{th} = 3.1$ A. That is, the loop will oscillate without hysteresis losses provided the current induced by the movement is smaller than $i_{th}$. This is actually the case, since the maximum value of the rectified current necessary for levitation is[2] $i_{max} = 2i_0 = 2mg/(aB_0) = 0.94$ A. We conclude that if the loop oscillates in a vacuum, the only remaining source of energy dissipation will be the viscous drag of FL oscillating inside their pinning wells in a surface layer of thickness $\lambda_{eff}$. According to [6], the power dissipated per unit volume is $P = \frac{1}{2} (\omega\eta)^2/((\omega\eta)^2 + k^2) J^2 (H/H_{c2})\rho_n$, where we neglect the FL mass and $\omega = \Omega$. Inserting the figures into the expressions for the viscosity and for the elastic constant one obtains $\eta = 3.29 \times 10^{-8}$ and $k = 1200$ ( all in MKS units), which makes the $\eta/k$ ratio extremely small. This leads to a simplification in the expression for $P$, which can be written as

$$P = \frac{1}{2} (\Omega\eta/k)^2 J^2 (H/H_{c2})\rho_n \qquad (2)$$

The currents in the loop flow within a surface layer of thickness $\lambda_{eff} = 0.63$ μm for $B_2 = 0.3$ T, and 0.89 μm, for $B_1 = 0.6$ T( we take the same $J_c$ for both fields), so that the effective cross-sectional area of wire penetrated by currents is $S_s =$ $2\pi r \lambda_{eff}$ = 1.27x10$^{-9}$ m$^2$ if the field is $B_2$, and 1.79x10$^{-9}$ m$^2$ if the field is $B_1$. The current density in (2) is obtained from the amplitude of the alternating part of the rectified current, that is, $J = i_0/S_s$, for each value of $S_s$. Therefore, the average value of $P$ taken around the loop is 1.2 x10$^{-6}$ W/m$^3$. Such power is dissipated in a thin tubular shell where the FL oscillations actually take place around the loop. Taking into consideration the variations in $\lambda_{eff}$, the effective power dissipated in the loop is the averaged product of $P$ times the volume of the tubular shells, so that $P_{eff}$ = 4.1x10$^{-16}$ W. The quality factor $Q$ is defined as the ratio $U\Omega/P_{eff}$, where $U$ is the total energy of the oscillating system. In this case $U = mgx_0 = Li_0^2$ = 3.24x10$^{-8}$ J, which results in $Q$ = 8.8x10$^{10}$. This result is one order of magnitude greater than the quality factor of a state-of-the-art optical microcavity described in [1].

3. Discussion and Conclusions.

This paper, together with [2], has described how a simple oscillator might be built to work at low frequencies and still have an extremely high quality factor. Such effect is a direct consequence of the negligible resistivity at low frequencies of a strong-pinning superconductor wire, and also of the possibility of adapting the operational conditions so that there are no hysteresis losses. Although we have mentioned in[2] the prospective application of the oscillator in energy storage, a novel and potentially important application as part of a measurement system might be devised. We note that if the probe of a superconducting quantum interference device (SQUID) is placed close to the center of the loop, the rectified magnetic field generated by the currents in the loop will be detected. The currents that produce such field are directly proportional to the static force acting upon the loop ( its own weight *mg*, in the example given in this paper). In view of the stability of the oscillator, minute variations of the loop weight might be measured by the SQUID with all its available accuracy. In the example given in this letter the field produced by the currents would be on the order of $10^{-3}$ Tesla ( $\approx \delta$). If the SQUID probe is a ring of 1 cm$^2$ area, its sensitivity to single flux quanta might permit the measurements of magnetic field variations of $\approx 10^{-11}$ T. That is, the field produced by the currents might be determined to a precision of 1 part in $10^8$. The same precision would therefore be associated to the knowledge of the parameter *g*. This implies the possibility of the immediate application of this oscillator to measurements of extremely small variations in the gravitational field, such as those expected from gravitational waves, for instance.

Acknowledgement: The author wishes to thank Professor Said Salem-Sugui for helpful discussions.

Figure caption.

Figure 1: A square type-II superconductor loop of side $l$ is subjected to a distribution of uniform static fields $B_1$ and $B_2$ greater than the lower critical field. Note that while the loop oscillates in the vertical direction no part of its wires is subjected to variable static magnetic fields, something essential to avoid hysteresis losses[2].

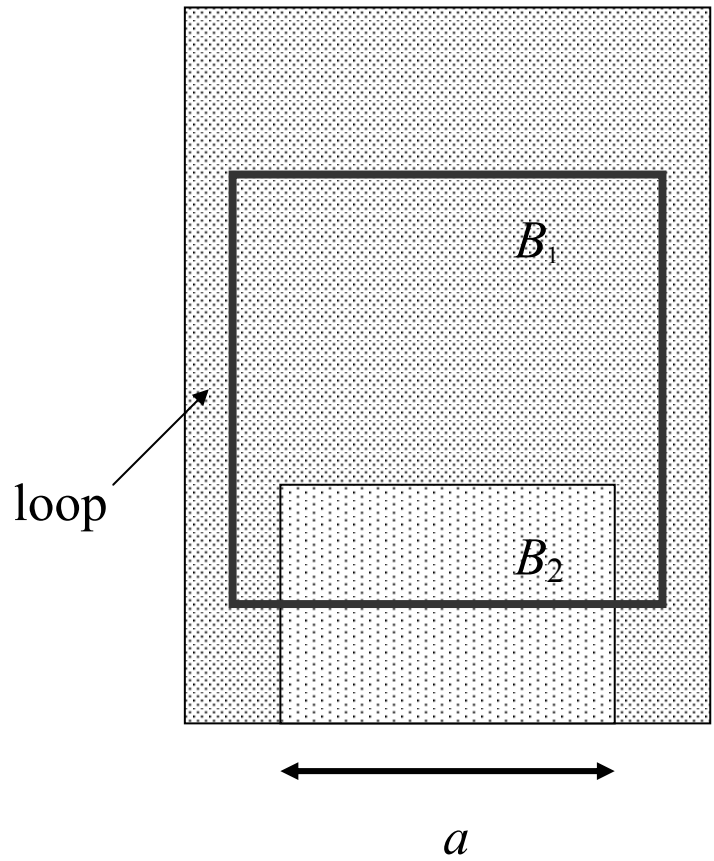

Figure 1